\documentstyle[a4,12pt,epsfig,amsmath]{article}
 \def\relic{\Omega_{DM
}}
\def\lsim{\raise0.3ex\hbox{$\;<$\kern-0.75em\raise-1.1ex\hbox{$\sim\;$}}}
\def\gsim{\raise0.3ex\hbox{$\;>$\kern-0.75em\raise-1.1ex\hbox{$\sim\;$}}}

\def\bsg{$b\to s\gamma$}

\def\asusy{a^{\rm SUSY}_\mu}

\DeclareMathAlphabet   {\mathsc}{OT1}{cmr}{m}{sc} 
 
\def\[{\left [} 
\def\]{\right ]} 
\def\({\left (} 
\def\){\right )}

\newcommand{\gappeq}{\mathrel{\rlap {\raise.5ex\hbox{$>$}} 
{\lower.5ex\hbox{$\sim$}}}} 
\newcommand{\lappeq}{\mathrel{\rlap{\raise.5ex\hbox{$<$}} 
{\lower.5ex\hbox{$\sim$}}}}

\newcommand{\bea}{\begin{eqnarray}}
\newcommand{\eea}{\end{eqnarray}}

\begin{document}

\pagestyle{empty}


\rightline{FTUAM 04/18}
\rightline{IFT-UAM/CSIC-04-43}
\rightline{hep-ph/0407352}
\rightline{July 2004}

\vspace{1.cm}

\begin{center}

{\Large {\bf A comparison between 
direct and indirect\\ dark matter search
}}
\vspace{1cm}\\
{\large Y. Mambrini 
{\small and} C. Mu\~noz
}
\vspace{1cm}\\
Departamento de F\'isica Te\'orica C-XI and Instituto de F\'isica
Te\'orica C--XVI,\\
Universidad Aut\'onoma de Madrid, Cantoblanco,
28049 Madrid, Spain.
\vspace{1cm}\\
 
\end{center}

\abstract{
We carry out a comparison between different kinds of
experiments for the detection of neutralino dark matter.
In particular, we compare the theoretical
predictions for the direct detection of these particles,
through their elastic scattering
on target nuclei through nuclear recoils,
with those for their indirect detection
through the gamma rays produced by their annihilation in the
galactic center. 
For the comparison we pay special attention
to points which will be testable by 
very sensitive experiments, such as e.g. EDELWEISS II and GENIUS 
for direct detection, 
and GLAST for indirect detection.
We perform the analysis in 
the framework of a general SUGRA theory, allowing the presence 
of non-universal
soft scalar and gaugino masses in the parameter space.  
As a conclusion, we find 
that indirect detection experiments will be able to test larger 
regions of the parameter space than direct detection ones.


}

\vspace{3cm}

\newpage

\tableofcontents

\vspace{3cm}


\setcounter{page}{1}
\pagestyle{plain}

\section{Introduction}




A
weakly interacting massive particle (WIMP)
is one of the most interesting candidates for dark matter in the Universe.
The main reason being that WIMPs can be left over from the
Big Bang in the right amount to account for the observed matter
density.
Let us recall that 
the observations  
of galactic rotation curves \cite{Persic}, cluster of
galaxies,
and large scale flows \cite{Freedman},
imply
$0.1\lsim \relic h^2\lsim 0.3$. In addition,
the recent data obtained by the WMAP satellite \cite{wmap03-1}
confirm that dark matter must be present with
$0.094\lsim\relic h^2 \lsim 0.129$.

Many direct 
detection
experiments have been carried out around the world in order to detect
WIMPs \cite{mireview}.
These particles 
cluster gravitationally in our galaxy,
and therefore there must be a flux of them on the Earth. 
This flux can be detected
by observing the elastic scattering of WIMPs on target nuclei through nuclear 
recoils.
In fact, one of the experiments, the DAMA collaboration, 
reported data favouring the existence of a WIMP signal \cite{dama}.
Taking into account uncertainties on the halo model,
it was claimed that 
the preferred range of the WIMP-nucleon cross section
is $\sigma \approx 10^{-6}-10^{-5}$ pb 
for a WIMP mass smaller than $500-900$ GeV \cite{halo,dama}.
Unlike this spectacular result other collaborations such as 
CDMS \cite{experimento2} and EDELWEISS \cite{edelweiss}, 
claim to have excluded important regions of the DAMA 
parameter space.

This controversy may be solve in the future by these and other
projected experiments.
For example,
GEDEON \cite{IGEX3}  
will be able to explore
positively a WIMP-nucleon 
cross section
$\sigma \gsim 3\times 10^{-8}$ pb. Similar cross sections
will also be tested by EDELWEIIS II.
CDMS Soudan (an expansion of the CDMS experiment
in the Soudan mine), 
will be able to test 
$\sigma \gsim 2\times 10^{-8}$ pb.
In fact, while this paper was in preparation, the Soudan experiment
reported its first results \cite{soudanmine},
and these are even stronger than those of EDELWEISS.
Let us finally mention that the 
most sensitive 
detector will be 
GENIUS \cite{HDMS2}, 
which will be able to test a WIMP-nucleon cross section
as low as $\sigma\approx 10^{-9}$ pb.

On the other hand, there are also promising methods for the
indirect detection of WIMPs through the analysis of their annihilation
products \cite{Feng}. One of them consists of 
detecting the gamma rays produced by
these annihilations in the galactic halo.
For this, one uses atmospheric Cherenkov telescopes or space-based
gamma-ray detectors. 
Similarly to the case of direct detection, one of the space-based experiments,
the Energetic Gamma-Ray Experiment Telescope (EGRET) 
on the Compton Gamma-Ray Observatory, 
detected a signal \cite{EGRET} that apparently cannot be explained with the 
usual gamma-ray background. In particular, after
five years of mapping the gamma-ray sky up to an energy
of about 20 GeV, there are evidences of a source of 
gamma rays above about 1 GeV in the galactic center region
with a value for the integrated flux $\Phi_{\gamma}$ of about
$10^{-8}$ cm$^{-2}$\ s$^{-1}$.
It is worth noticing, however, that alternative explanations
for this result have been proposed 
modifying conveniently  
the standard theory of galactic gamma ray \cite{modi}.

As in the case of direct detection, projected experiments might
clarify the situation. For example,
a larger sensitivity will be provided by the Gamma-ray Large Area
Space Telescope (GLAST) \cite{GLAST}, which is scheduled for launch
in 2006. 
In addition, energies
as high as 300 GeV can be measured testing models with heavy 
dark matter more
easily than EGRET.
Even if alternative explanations arise for the events observed
by EGRET, 
GLAST will be able to detect a smaller flux of gamma rays from
dark matter $\Phi_{\gamma}\sim10^{-10}$ cm$^{-2}$\ s$^{-1}$.
Besides,
ground-based atmospheric Cherenkov telescopes
are complementary to space-based detectors.
For example, the High Energy Stereoscopic System (HESS) 
experiment \cite{HESS}, which has begun operations with four
telescopes\footnote{After this paper was sent for publication,
the first data from HESS appeared \cite{HESS2}. Although they observe
gamma rays in a particular direction, these would correspond
to a neutralino heavier than 12 TeV, a range disfavoured
by particle physics where supersymmetric particules should not be
heavier than about 1 TeV. 
Thus, an astrophysical source is a more
likely explanation for these data.}, 
has a large effective area of $~10^{5}$  m$^2$ (versus 1 m$^2$ for GLAST) 
allowing the detection of very high energy fluxes $\gsim 1$ TeV.
On the contrary, the energy threshold is larger, $E_{\gamma}> 60$ GeV.

Given the experimental situation described above, concerning the
detection of dark matter, it is natural to wonder how big 
the scattering (annihilation) cross section for its direct (indirect) 
detection can be.
Obviously,
this analysis is crucial in order to know the
possibility of detecting dark matter 
in the experiments. However,
the answer to this question depends on the particular
WIMP considered.
The leading candidate in this class of particles is the
lightest neutralino, $\tilde{\chi}^0_1$,
which appears in supersymmetric (SUSY) extensions
of the standard model \cite{Goldberg}.

For the case of direct detection, the analysis of neutralino dark matter
has been carried out with the
following results. 
In the usual minimal supergravity (mSUGRA) scenario, where the soft terms
of the minimal supersymmetric standard model (MSSM) 
are assumed to be universal at the unification scale, 
$M_{GUT} \approx 2\times 10^{16}$ GeV,
and radiative electroweak symmetry
breaking is imposed, 
the cross section turns out
to be small, $\sigma_{\tilde{\chi}_1^0-p}\lsim 3\times 10^{-8}$ pb \cite{mireview}.
Clearly, in this case, current direct-detection 
experiments are not sufficient and
more sensitive detectors
producing further data, such as e.g. 
GENIUS, 
are needed.
However, this result can be modified by taking into account
possible departures from the mSUGRA scenario.
In particular,
a more general situation in the context
of SUGRA than universality, the presence of non-universal
soft scalar and gaugino masses, may increase the cross section 
significantly 
(for recent works see e.g. \cite{cggm03-1,muchos,cggm03-12} and references
therein).

For the indirect detection of dark matter through gamma rays,
similar theoretical analyses have been carried out for
the case of mSUGRA 
 \cite{Feng,Ullio,Boer,Wang,Bin2,farrill,mammu}.
Neutralino annihilations in the galactic center region 
seem not to be able
to explain the excess of gamma-ray fluxes detected by EGRET.
In fact, 
only
small regions of the
parameter space of mSUGRA may be compatible with the sensitivity of
GLAST
and HESS, for standard dark matter density profiles such as the
Navarro, Frenk and White (NFW) one.
However, as in the direct detection case, when non-universal
scalar and gaugino masses are allowed,
the annihilation cross section may increase  
significantly, producing larger gamma-ray fluxes \cite{mammu}\footnote{For
analyses in the context of the effMSSM scenario,
where the parameters are defined directly at the electroweak scale,
see e.g. the recent works \cite{hooper,Bottino,Profumo}.}.

Given the above theoretical results, it would be interesting
to carry out a comparison of the sensitivity of 
both kinds of experiments, namely direct versus indirect
detection of dark matter.
This is precisely the aim of this paper.
In Section~2
we will review very briefly the above results
concerning the direct and indirect detection of neutralino
dark matter in a general SUGRA theory.
Taking into account this discussion,
in Section~3 we will compare the sensitivity of both kinds of
experiments in the parameter space of SUGRA.
In particular, we will pay special attention
to points which will be testable by very sensitive experiments,
such as e.g. EDELWEISS II, GENIUS, GLAST.
Finally, the conclusions are left for Section~4.



\section{SUGRA predictions for the detection of dark matter}

In this section we will briefly review 
the SUGRA scenario and its impact on the direct and indirect
detection of neutralino dark matter. 
Let us first recall that in mSUGRA,
where the soft terms 
are assumed to be universal,
and radiative electroweak symmetry
breaking is imposed, 
one has only four free parameters defined at the GUT scale:
the soft scalar mass $m$, the soft gaugino mass $M$, 
the soft trilinear coupling $A$, and 
$\tan\beta$.
In addition, the sign of the Higgsino mass parameter, $\mu$,
remains also undetermined by the 
minimization of the Higgs potential.

Using these parameters the neutralino-proton cross section has been 
analysed exhaustively in the literature. The result is that
the scalar cross section is small, 
and only future detectors such as GENIUS will be 
able to cover some regions of the parameter space 
(for a recent analysis, see e.g. \cite{cggm03-1}).
Similar results are obtained when gamma-ray fluxes from the
galactic center are studied in the framework of mSUGRA. 
In particular, the annihilation cross section
of neutralinos is small, and we need
future detectors such as GLAST to analysed at least some regions.

However, as discussed in detail in \cite{cggm03-1,muchos,cggm03-12},
in the context of direct detection, and in \cite{mammu}
for indirect detection, both cross sections mentioned above
can be increased significantly when the structure of mSUGRA
soft terms is abandoned\footnote{
In this sense, it is worth noticing that
such a non-universal structure 
can be recovered in the low-energy limit of some
phenomenologically appealing string
scenarios.
This is the case, 
for example, 
of 
D-brane constructions in the type I 
string \cite{Rigolin}. An analysis of direct detection of neutralinos 
in these scenarios was carried out in \cite{nosotros}.
An analysis of the gamma-ray fluxes might also be interesting \cite{us}.
On the other hand, AMSB scenarios in an effective heterotic framework
were studied in Ref.~\cite{Bin2,Mambrini}.
}.
Therefore larger regions of the parameter space can be covered
by the experiments.
Following those works, 
there are several interesting possibilities when non-universal
scalars and gauginos are considered in a general SUGRA theory.

On the one hand, 
parameterising the non-universalities in the Higgs sector, at the
GUT scale, as follows:
\begin{equation}
  m_{H_{d}}^2=m^{2}(1+\delta_{1})\ , \quad m_{H_{u}}^{2}=m^{2}
  (1+ \delta_{2})\ ,
  \label{Higgsespara}
\end{equation}
one has three representative cases:
\begin{eqnarray}
a)\,\, \delta_{1}&=&0\ \,\,\,\,\,\,\,\,,\,\,\,\, \delta_2\ =\ 1\ ,
\nonumber\\
b)\,\, \delta_{1}&=&-1\ \,\,\,\, ,\,\,\,\, \delta_2\ =\ 0\ ,
\nonumber\\
c)\,\, \delta_{1}&=&-1\ \,\,\,\, ,\,\,\, \delta_2\ =\ 1 
\ .
\label{3cases}
\end{eqnarray}
On the other hand, parameterising the non-universalities in the
gaugino sector as follows:
\begin{eqnarray}
  M_1=M\ , \quad M_2=M(1+ \delta'_{2})\ ,
  \quad M_3=M(1+ \delta'_{3})
  \ ,
  \label{gauginospara}
\end{eqnarray}
where $M_{1,2,3}$ are the bino, wino and gluino masses, respectively,
one has the representative case:
\begin{eqnarray}
d)\,\, \delta'_{3}&=&-0.4\ \,\,\,\,\,\,\,\,,\,\,\,\, \delta'_2\ =\ 0\ .
\label{1cases}
\end{eqnarray}

Of course, as discussed in detail in \cite{cggm03-12},
the combination of both non-universalities, scalars and gauginos,
gives also rise to other interesting possibilities. 
In particular, these are choices giving rise to very
light neutralinos. These are obtained for example 
using the  scalar
non-universalities given in Eq.~(\ref{3cases}), 
and adding gaugino non-universalities.
A representative case is $\delta'_{2,3}=3$

\begin{figure}[!]
    \begin{center}
\centerline{
       \epsfig{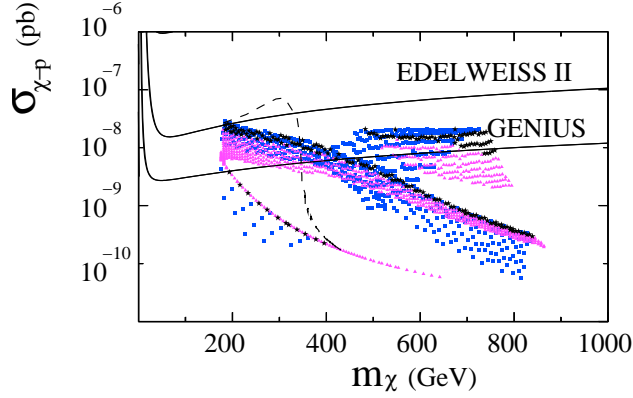}\hskip 1cm
       }
          \caption{{\footnotesize
Scatter plot of 
the scalar neutralino-proton cross section
$\sigma_{\tilde{\chi}_1^0-p}$ 
as a function of 
the neutralino mass 
$m_{\tilde{\chi}_1^0}$ in the 
non-universal scalar case {\it a)} 
$\delta_1=0$, $\delta_2=1$, discussed in Eq.~(\ref{3cases}),
for
$\tan\beta=35$ and $A=0$,
with $\mu>0$.
The region to the right of the 
dashed line
corresponds to
$\asusy< 7.1\times10^{-10}$, and would be excluded (allowed) by
electron
(tau) data.
The points in the figure 
fulfil all other experimental constraints and they have 
$0.129<\Omega_{\tilde{\chi}^0_1}h^2<0.3$ 
(magenta triangles),
$0.094<\Omega_{\tilde{\chi}^0_1} h^2<0.129$ (black stars),
and
$0.03<\Omega_{\tilde{\chi}^0_1} h^2<0.094$ (blue boxes) 
with the appropriate rescaling of the density of neutralinos
in the galaxy as discussed in the text. 
>From top to bottom, 
the upper area bounded by the first (second) solid line  
will be analized by the projected EDELWEISS II (GENIUS) experiment.
}}
        \label{fig:nonuniv1tb35rapportt}
    \end{center}

\end{figure}

\begin{figure}
   \begin{center}
\centerline{
       \epsfig{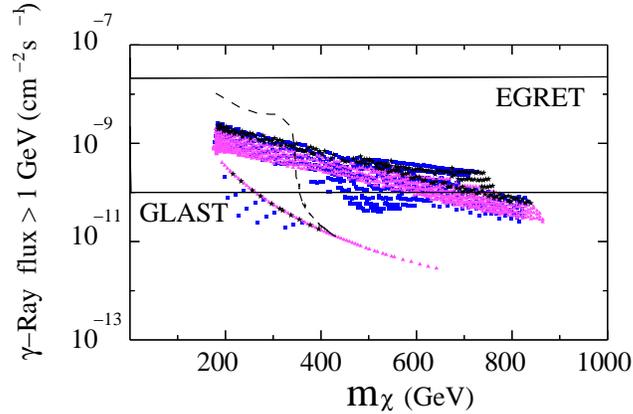}\hskip 1cm
       }
          \caption{{\footnotesize 
The same as in Fig.~\ref{fig:nonuniv1tb35rapportt}
but for 
gamma-ray integrated flux $\Phi_{\gamma}$ 
for a threshold of 1 GeV
versus the neutralino mass
$m_{\tilde{\chi}_1^0}$.
A NFW profile is used.
Now,
from top to bottom, 
the first solid line  
corresponds to the signal reported by EGRET, and
the upper area bounded by the second solid line will be analyzed by 
GLAST experiment.
}}
        \label{fig:nonuniv1tb35flux}
    \end{center}
\end{figure}

As mentioned above, one can find 
in \cite{cggm03-1,cggm03-12}, and \cite{mammu},
a detailed study of all these cases, and in particular
of 
the computation of neutralino-proton cross sections  $\sigma_{\tilde{\chi}_{1}^{0}-p}$
and gamma-ray fluxes $\Phi_{\gamma}$, respectively.
As an example, we show in Fig.~\ref{fig:nonuniv1tb35rapportt}
the values of $\sigma_{\tilde{\chi}_{1}^{0}-p}$
as a function of the neutralino mass
$m_{\tilde{\chi}_1^0}$, for the case
{\it a)}
with $\tan\beta=35$, $A=0$ and $\mu>0$.
The areas that will be 
covered in the future by two very sensitive detectors, EDELWEISS II and GENIUS,
are shown.
Likewise, for the same case, we show in Fig.~\ref{fig:nonuniv1tb35flux}
the values of 
$\Phi_{\gamma}$ as a function of 
$m_{\tilde{\chi}_1^0}$.
In addition to EGRET, we show in the figure the region that will
be covered by GLAST experiment.
For the sake of definiteness, we use in our analysis of the gamma-ray
flux an angular acceptance of the detector
$\Delta \Omega=10^{-3}$ sr,
and a NFW profile \cite{Navarro:1996he}.
The resuls
can easily be extended to other profiles.
For example, 
the flux produced by
the Moore et al. profile \cite{Moore:1999gc}
is about two orders of magnitude larger than the one produced
by the NFW, and this is about two orders of magnitude larger
than the flux produced by the Isothermal and Kravtsov et 
al. \cite{Kravtsov:1997dp} 
profiles 
(see e.g. Table~1 in \cite{mammu}).

Let us remark that in the figures,
only points allowed by all experimental constraints are shown.
In particular, the lower
bounds on the masses of the SUSY particles and on the
lightest Higgs have been implemented, as well as the experimental
bounds on the branching ratio of the \bsg\ process and on 
$\asusy$. 
Concerning the latter, we have taken into account the
recent experimental result for the muon
anomalous magnetic moment \cite{g-2}, as well as the most recent
theoretical evaluations of the Standard Model contributions
\cite{newg2}. It is found that when $e^+e^-$ data
are used the experimental excess in $(g_\mu-2)$ would constrain a
possible SUSY contribution to be
$\asusy=(27.1\,\pm\,10)\times10^{-10}$. 
At $2\sigma$ level this implies 
$7.1\times10^{-10}\lsim\asusy\lsim47.1\times10^{-10}$, and therefore
only cases with $\mu>0$ are interesting. 
It is worth noticing here that when tau data are used a smaller
discrepancy with the experimental measurement is found.
In order not to exclude the latter possibility we will 
show the line $\asusy= 7.1\times10^{-10}$ in the figures throughout the paper. 
The region of the parameter space to the right of this line
will be forbidden (allowed) if one consider electron (tau) data.

Concerning the astrophysical bounds on the relic density, we show
in the figures 
with black stars
points with  $0.094<\Omega_{\tilde{\chi}^0_1} h^2<0.129$, and therefore favoured
by WMAP.
Points with $0.129<\Omega_{\tilde{\chi}^0_1} h^2<0.3$ are shown with magenta
triangles.
We also consider the possibility that not all the dark matter is made
of
neutralinos, allowing
$0.03<\Omega_{\tilde{\chi}^0_1} h^2<0.094$. In this case we have 
to rescale the density of 
neutralinos in the galaxy $\rho(r)$ 
by a factor
$\Omega_{\tilde{\chi}^0_1}h^2/ 0.094$. 
Points 
corresponding to this possibility are
shown with blue boxes.
For details on how the rest of the experimental bounds are implemented,
and which programs are used to carry out the evaluation, 
see \cite{mammu}.

\newpage

\section{A comparison between direct and indirect detection}

As discussed in detail in previous sections
there are many experiments in preparation for the direct and
indirect detection
of dark matter, and they will be able to cover an 
important range of the parameter space of SUSY models.
Thus, in principle, it might be possible 
that this means of detection will be able to probe a dark matter signal 
before the end of the decade. Therefore, it becomes 
interesting to compare both kinds of experiments.
In particular, we will compare the direct detection of neutralinos,
through their elastic scattering on target nuclei through nuclear 
recoils,
with the indirect detection of neutralinos,
through the gamma rays produced by their annihilation in the
galactic center.



\begin{figure}[!]
    \begin{center}
\centerline{
       \epsfig{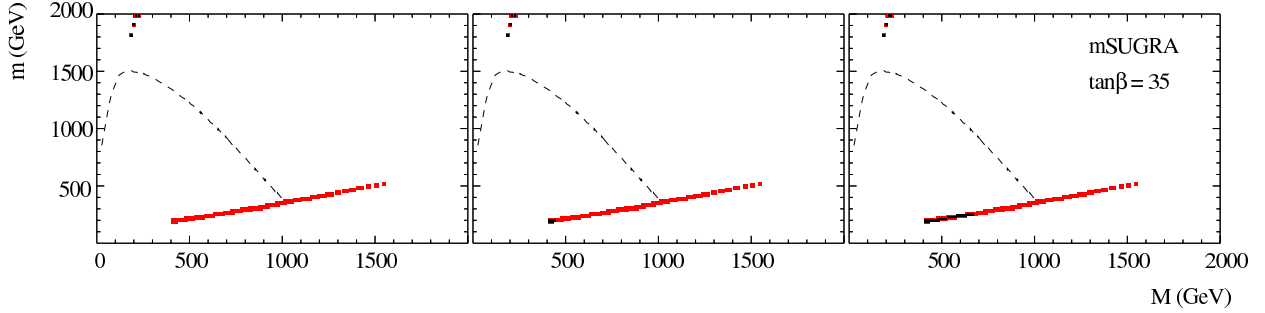}
       }
          \caption{{\footnotesize 
Points fulfilling
all experimental constraints with
$0.1<\Omega_{\tilde{\chi}^0_1}h^2<0.3$ 
(red) in the parameter space of the mSUGRA scenario
($m$, $M$) for $\tan\beta=35$, $A=0$ and $\mu>0$.
The black colour on top of these indicates those points
that will be testable by EDELWEISS II (left),
GENIUS (center), and GLAST (right) assuming the
NFW density profile in the theoretical compution.
The region to the right of the 
dashed line
corresponds to
$\asusy< 7.1\times10^{-10}$, and would be excluded (allowed) by
electron
(tau) data.
 }}
        \label{fig:univ1tb35scan}
    \end{center}
\end{figure}

Let us start with the simple mSUGRA scenario where the soft
terms are universal at the GUT scale.
In Fig.~\ref{fig:univ1tb35scan}  
we show in red the points fulfilling
all experimental and astrophysical constraints 
discussed in the previous section, in the parameter space
($m$, $M$).
The black colour on top of these indicates those points
that will be testable by the experiments.
In particular, on the left and central frames we use
two very sensitive experiments for the direct detection 
of dark matter, EDELWEISS II and GENIUS, respectively.
On the right frame we use the indirect detection experiment GLAST,
assuming the NFW density profile for the theoretical computation
of the signal.
This figure is for $\tan\beta=35$, and throughout the paper we are
using $A=0$ and $\mu>0$.

\begin{figure}[!]
    \begin{center}
\centerline{
       \epsfig{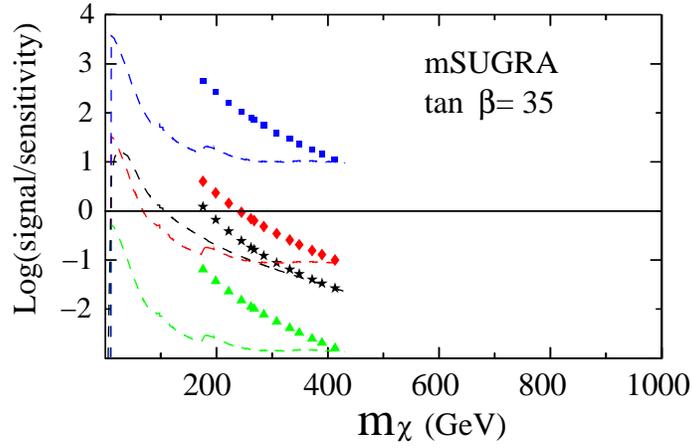}
       }
          \caption{{\footnotesize 
Log($\frac{signal}{sensitivity}$) as a function of 
the neutralino mass $m_{\tilde{\chi}_1^0}$ for points
fulfilling
all experimental constraints with
$0.094<\Omega_{\tilde{\chi}^0_1}h^2<0.129$.
The solid line separates points which are detectable (upper region)
from those which are not (lower region).
Black stars are associated to the case of direct detection, where
the signal corresponds to $\sigma_{\tilde{\chi}_1^0-p}$ and the sensitivity used
is the one of GENIUS.
The rest of the points are associated to the case of indirect detection
through gamma-rays, where the signal corresponds to 
$\Phi_{\gamma}$ and the sensitivity used is the one of GLAST.
Results for different density profiles are shown: 
Moore et al. (blue boxes), NFW (red diamond), Kavtsov et al. 
(green triangle). For each profile,
the region to the right of the 
dashed line
corresponds to
$\asusy< 7.1\times10^{-10}$, and would be excluded (allowed) by
electron
(tau)
data.
}}
        \label{fig:univ1tb35arapport}
    \end{center}

\end{figure}

Although indirect detection experiments will be able to test
larger regions of the parameter space than direct detection
ones, these regions are in any case small, and most of the points
will avoid detection.
Let us recall, however, that we are working with a NFW profile
in our computation of gamma-ray fluxes.
Since the Moore et al. profile is about two orders of magnitude
larger, one should expect more detectability.
This is summarized for different density profiles
in Fig.~\ref{fig:univ1tb35arapport},
where Log($\frac{signal}{sensitivity}$) as a function of 
$m_{\tilde{\chi}_1^0}$ is shown.
For direct detection the signal consists of
$\sigma_{\tilde{\chi}_1^0-p}$ and the sensitivity used in the plot 
is the one of the GENIUS
experiment (see Fig.~\ref{fig:nonuniv1tb35rapportt}).
For indirect detection the signal consists of
$\Phi_{\gamma}$ and the sensitivity used is the one of the
GLAST experiment (see Fig.~\ref{fig:nonuniv1tb35flux}). 
The solid line separates points which are detectable (upper region)
from those which are not (lower region).

We have also analyzed very large values of $\tan\beta$, such as 50.
In this case the difference of sensitivity between direct and 
indirect detection experiments is more obvious. 
Whereas the region of the parameter space that will be tested by
GENIUS is very similar to the one of the previous case with
$\tan\beta=35$,
the region tested by 
GLAST increases significantly
Note, finally, that EDELWEISS II will not be able to detect any neutralino in
the framework of mSUGRA.

\begin{figure}[!]
    \begin{center}
\centerline{
       \epsfig{file=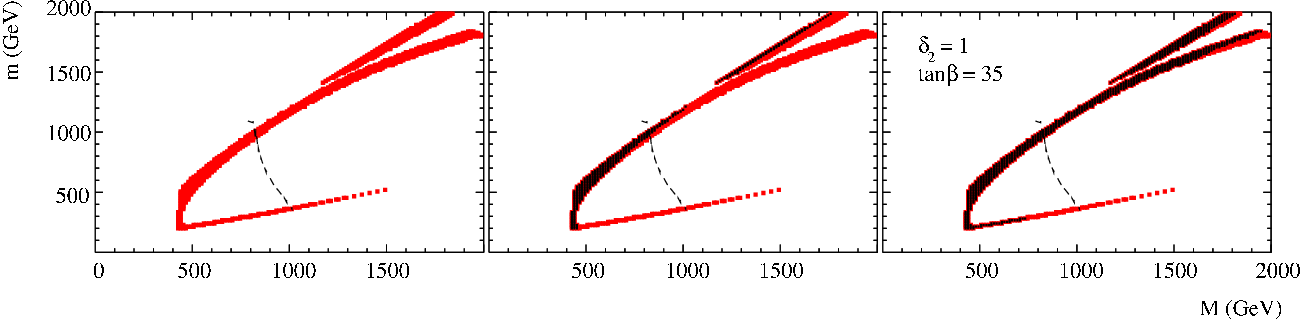,width=1.\textwidth}
       }
          \caption{{\footnotesize 
The same as in Fig.~\ref{fig:univ1tb35scan} but for the
non-universal scalar case {\it a)} 
$\delta_1=0$, $\delta_2=1$, discussed in Eq.~(\ref{3cases}).
 }}
        \label{fig:nonuniv1tb35scana}
    \end{center}
\end{figure}

\begin{figure}[!]
    \begin{center}
\centerline{
       \epsfig{file=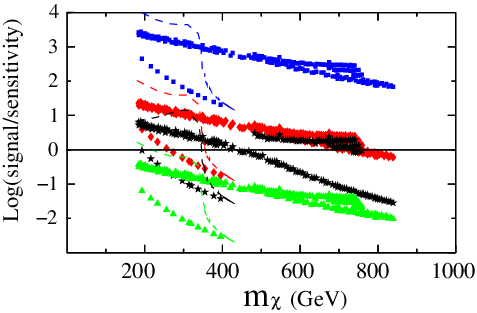,width=0.5\textwidth}
       }
          \caption{{\footnotesize 
The same as in Fig.~\ref{fig:univ1tb35arapport}
but for the
non-universal scalar case {\it a)} 
$\delta_1=0$, $\delta_2=1$, discussed in Eq.~(\ref{3cases}).
}}
        \label{fig:nonuniv1tb35arapporta}
    \end{center}

\end{figure}

\begin{figure}[!]
    \begin{center}
\centerline{
       \epsfig{file=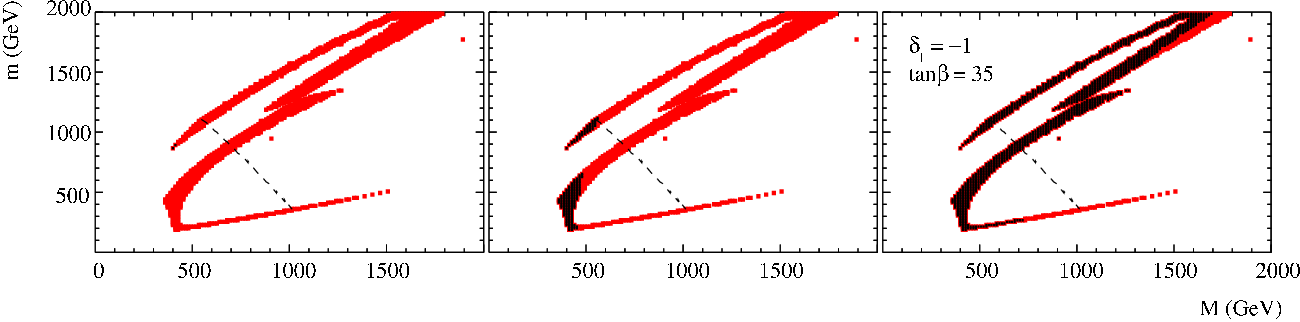,width=1.\textwidth}
       }
          \caption{{\footnotesize 
The same as in Fig.~\ref{fig:univ1tb35scan} but for the
non-universal scalar case {\it b)} 
$\delta_1=-1$, $\delta_2=0$, discussed in Eq.~(\ref{3cases}).
}}
        \label{fig:nonuniv1tb35scanb}
    \end{center}
\end{figure}

\begin{figure}[!]
    \begin{center}
\centerline{
       \epsfig{file=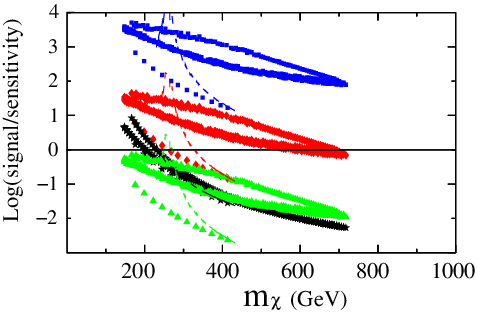,width=0.5\textwidth}
       }
          \caption{{\footnotesize 
The same as in Fig.~\ref{fig:univ1tb35arapport}
but for the
non-universal scalar case {\it b)} 
$\delta_1=-1$, $\delta_2=0$, discussed in Eq.~(\ref{3cases}).
}}
        \label{fig:nonuniv1tb35arapportb}
    \end{center}

\end{figure}

\begin{figure}[!]
    \begin{center}
\centerline{
       \epsfig{file=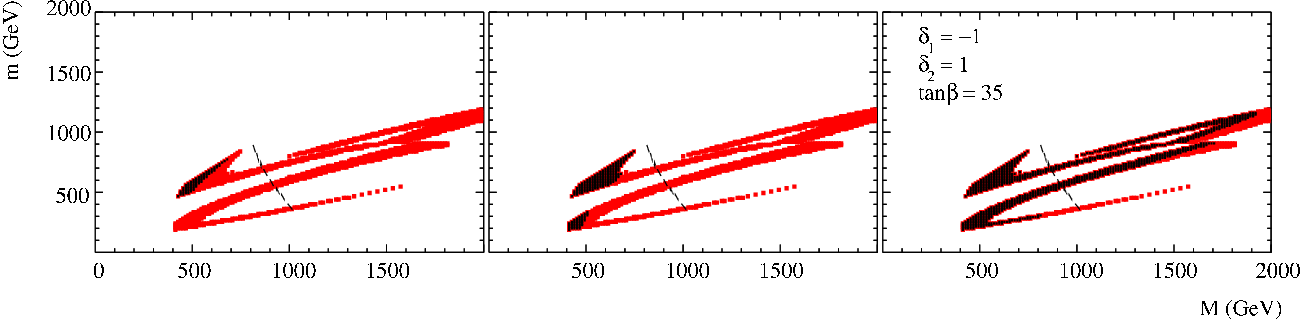,width=1.\textwidth}
       }
          \caption{{\footnotesize
The same as in Fig.~\ref{fig:univ1tb35scan} but for the
non-universal scalar case {\it c)} 
$\delta_1=-1$, $\delta_2=1$, discussed in Eq.~(\ref{3cases}).
}}
        \label{fig:nonuniv1tb35scanc}
    \end{center}
\end{figure}

\begin{figure}
    \begin{center}
\centerline{
       \epsfig{file=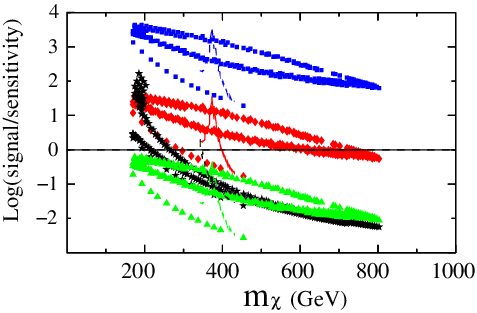,width=0.5\textwidth}
       }
          \caption{{\footnotesize 
The same as in Fig.~\ref{fig:univ1tb35arapport}
but for the
non-universal scalar case {\it c)} 
$\delta_1=-1$, $\delta_2=1$, discussed in Eq.~(\ref{3cases}).
}}
        \label{fig:nonuniv1tb35arapportc}
    \end{center}

\end{figure}

On the other hand, for non-universal Higgs masses the experimental situation
changes significantly. Larger regions of the parameter space
become available, mainly for GENIUS and GLAST. This 
can be seen in 
Figs.~\ref{fig:nonuniv1tb35scana} to
\ref{fig:nonuniv1tb35arapportc}, 
which can be compared with
Figs.~\ref{fig:univ1tb35scan}  and \ref{fig:univ1tb35arapport}
of mSUGRA.
This effect is clearly stronger for the indirect detection.
For example, only small regions of the parameter space
will be tested by GENIUS for cases
{\it b)} and {\it c)}, whereas
most of parameter space 
will be tested by GLAST.

\begin{figure}
    \begin{center}
\centerline{
       \epsfig{file=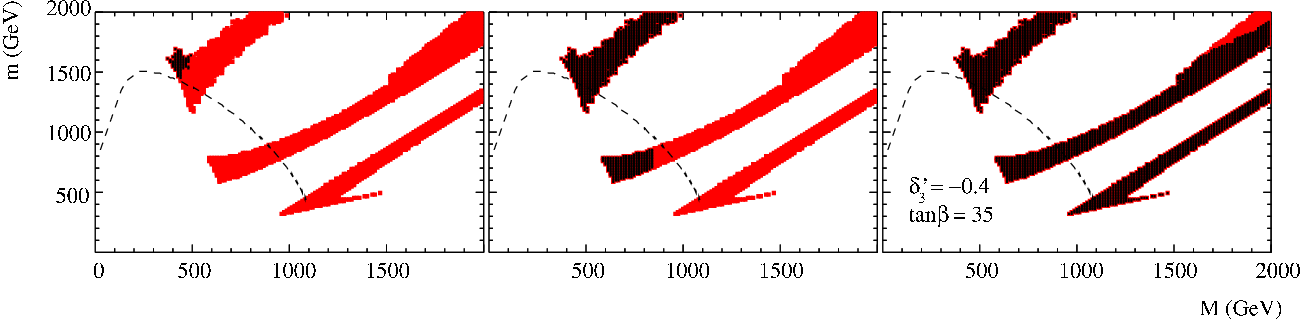,width=1.\textwidth}
       }
          \caption{{\footnotesize
The same as in Fig.~\ref{fig:univ1tb35scan} but for the
non-universal gaugino case {\it d)} 
$\delta'_2=0$, $\delta'_3=-0.4$, discussed in Eq.~(\ref{1cases}).
}}
        \label{fig:nonuniv1tb35scang}
    \end{center}

\end{figure}

\begin{figure}
    \begin{center}
\centerline{
       \epsfig{file=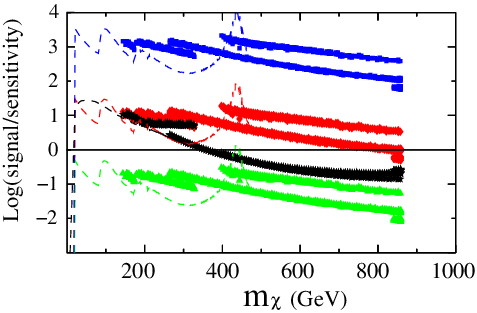,width=0.5\textwidth}
       }
          \caption{{\footnotesize 
The same as in Fig.~\ref{fig:univ1tb35arapport}
but for the
non-universal gaugino case {\it d)} 
$\delta'_2=0$, $\delta'_3=-0.4$, discussed in Eq.~(\ref{1cases}).
}}
        \label{fig:nonuniv1tb35arapportd}
    \end{center}

\end{figure}

The situation for non-universal gauginos
is also qualitatively similar, and we show this
in Figs.~\ref{fig:nonuniv1tb35scang} and 
\ref{fig:nonuniv1tb35arapportd}

\begin{figure}
    \begin{center}
\centerline{
       \epsfig{file=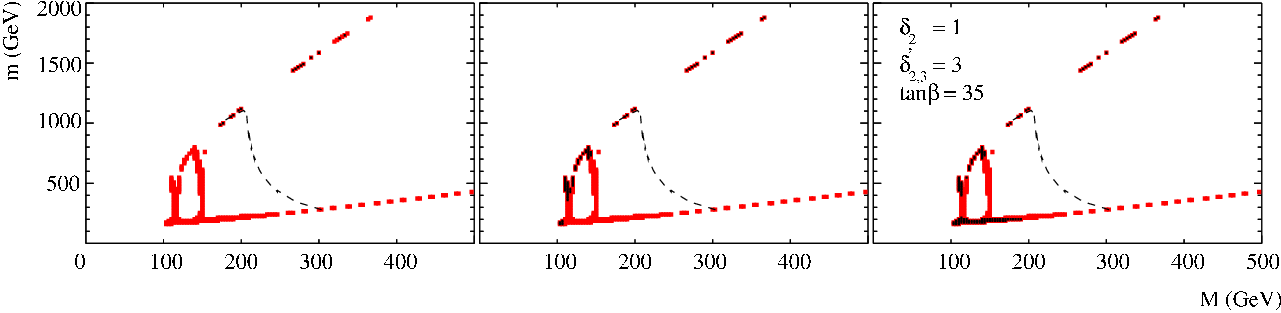,width=1.\textwidth}
       }
          \caption{{\footnotesize
The same as in Fig.~\ref{fig:univ1tb35scan} but for the
non-universal scalar and gaugino case  
$\delta_1=0$, $\delta_2=1$, $\delta'_2=\delta'_3=3$.
}}
        \label{fig:nonuniv1tb35scanag}
    \end{center}
\end{figure}

\begin{figure}
    \begin{center}
\centerline{
       \epsfig{file=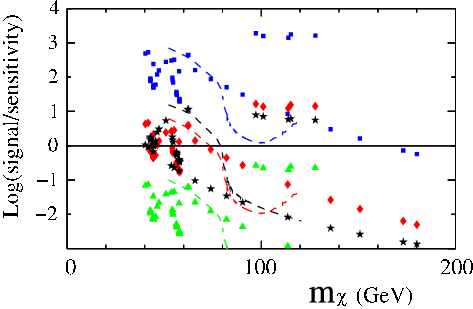,width=0.5\textwidth}
       }
          \caption{{\footnotesize 
The same as in Fig.~\ref{fig:univ1tb35arapport}
but for the non-universal scalar and gaugino case  
$\delta_1=0$, $\delta_2=1$, $\delta'_2=\delta'_3=3$.
}}
        \label{fig:nonuniv1tb35rapportag}
    \end{center}

\end{figure}

\begin{figure}
    \begin{center}
\centerline{
       \epsfig{file=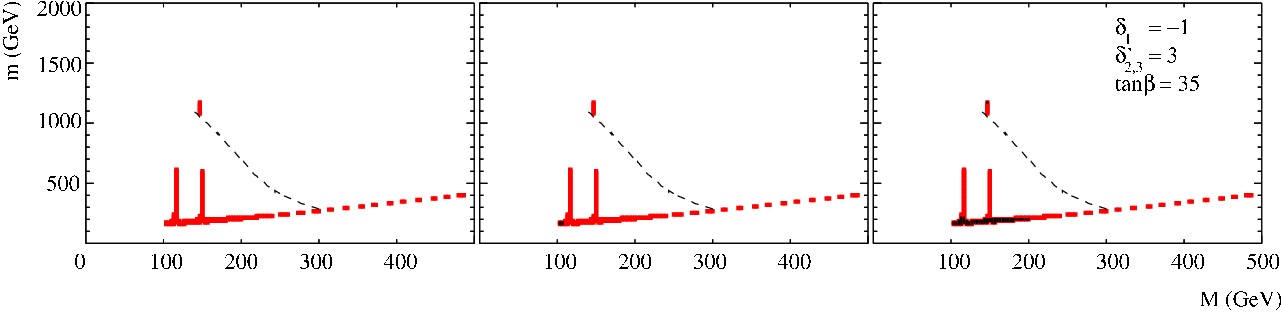,width=1.\textwidth}
       }
          \caption{{\footnotesize
The same as in Fig.~\ref{fig:univ1tb35scan} but for the
non-universal scalar and gaugino case  
$\delta_1=-1$, $\delta_2=0$, $\delta'_2=\delta'_3=3$.
}}
        \label{fig:nonuniv1tb35scanbg}
    \end{center}
\end{figure}

\begin{figure}
    \begin{center}
\centerline{
       \epsfig{file=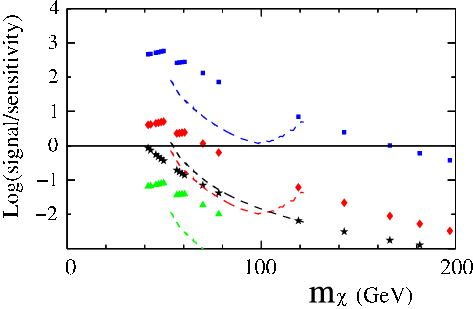,width=0.5\textwidth}
       }
          \caption{{\footnotesize 
The same as in Fig.~\ref{fig:univ1tb35arapport}
but for the non-universal scalar and gaugino case  
$\delta_1=-1$, $\delta_2=0$, $\delta'_2=\delta'_3=3$.
}}
        \label{fig:nonuniv1tb35rapportbg}
    \end{center}

\end{figure}

\begin{figure}
    \begin{center}
\centerline{
       \epsfig{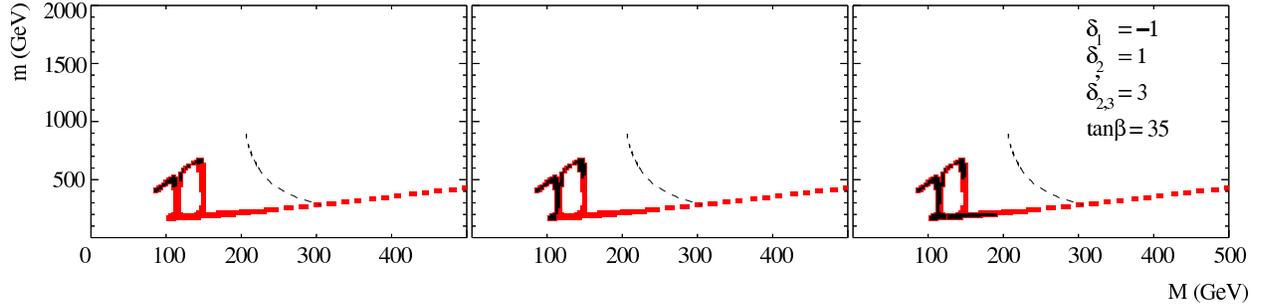}
       }
          \caption{{\footnotesize
The same as in Fig.~\ref{fig:univ1tb35scan} but for the
non-universal scalar and gaugino case  
$\delta_1=-1$, $\delta_2=1$, $\delta'_2=\delta'_3=3$.
}}
        \label{fig:nonuniv1tb35scancg}
    \end{center}
\end{figure}

\begin{figure}
    \begin{center}
\centerline{
       \epsfig{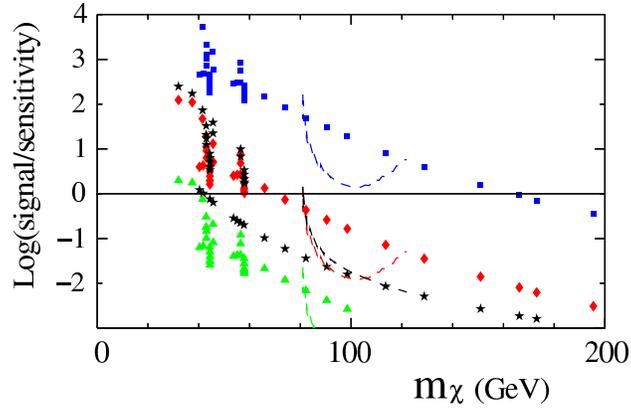}
       }
          \caption{{\footnotesize 
The same as in Fig.~\ref{fig:univ1tb35arapport}
but for the non-universal scalar and gaugino case  
$\delta_1=-1$, $\delta_2=1$, $\delta'_2=\delta'_3=3$.}}
        \label{fig:nonuniv1tb35rapportcg}
    \end{center}

\end{figure}

Finally, the other three representative cases mentioned in the previous
section,
with a very light neutralino mass,
can be seen in
Figs.~\ref{fig:nonuniv1tb35scanag} to 
\ref{fig:nonuniv1tb35rapportcg}.
The same comments as above, for non-universal scalars, 
can be applied here.
\newpage

\newpage








\newpage


\vskip 10cm


\newpage


\newpage



\newpage



\newpage


\newpage

\clearpage

\section{Conclusions}

There are two kinds of experiments in order to detect dark matter.
Direct detection experiments 
through the elastic scattering of dark matter particles
on target nuclei through nuclear 
recoils, and
indirect detection experiments through the analysis of dark matter
annihilation
products. 
We have compared both kinds of experiments for neutralino dark matter, 
paying special attention
in the indirect detection case to the gamma-ray fluxes produced by
the annihilation of neutralinos in the galactic center.

The discussion has been carried out using 
the parameter space of a general SUGRA theory, where not only
universal soft scalar and gaugino masses are allowed,
but also non-universalities may be present in both sectors, increasing
significantly the theoretical cross sections.
For the comparison we pay special attention
to points which will be testable by 
very sensitive experiments,
such as e.g. EDELWEISS II and GENIUS 
for direct detection, 
and GLAST for indirect detection.

In particular, 
we have determined the regions of the parameter space
that will  be testable by  
those two kinds of experiments and compare
between themselves.
We can summarize the results as follows.
Interesting direct detection experiments, 
such as e.g. EDELWEISS II, will only be able to cover
a extremely small region of the parameter space.
More powerful detectors such as GENIUS will cover larger regions,
but still not too large
compared with the parameter space of general SUGRA.
On the contrary, indirect detection experiments will be able to 
test important regions of the parameter space of SUGRA.

Of course, a signal in a direct detection experiment
would be a definitive proof of the existence of dark matter. 
However, we should keep in mind
that this might be difficult, depending on the region of the
parameter space where the neutralino dark matter is attached.
Therefore, at the end of the day, indirect detection experiments might
play a crucial role.


\vspace*{1cm}
\noindent{\bf Acknowledgements}

The work of Y. Mambrini was supported 
by the European Union under contract
HPRN-CT-2000-00148.
The work of C. Mu\~noz was supported
in part by the Spanish DGI of the
MEC 
under contracts BFM2003-01266 and FPA2003-04597;
and the European Union under contract 
HPRN-CT-2000-00148.

\newpage


\nocite{}
\bibliography{bmn}
\bibliographystyle{unsrt}

\end{document}